\documentclass[aps,a4paper,superscriptaddress,preprintnumbers,showpacs,amsmath,amssymb]{revtex4}
\usepackage{graphicx}
\usepackage{dcolumn}
\usepackage{color}
\usepackage{latexsym,amsfonts}
\usepackage{textcomp}
\usepackage{bm}

\usepackage{ulem}

\begin{document}

 \title{Comment on \\``On the implementation of CVC in weak
   charged-current proton-neutron transitions'' by  C. Giunti, arXiv:
   1602.00215 [hep--ph]}
 
 \author{A. N. Ivanov}\email{ivanov@kph.tuwien.ac.at}
 \affiliation{Atominstitut, Technische Universit\"at Wien,
   Stadionallee 2, A-1020 Wien, Austria}

\begin{abstract}
We show that the term maintaining conservation of the charged vector
current for the transitions ``neutron $\longleftrightarrow$ proton''
even for different masses of the neutron and proton (see T. Leitner
{\it et al.}, Phys. Rev. C {\bf 73}, 065502 (2006) and A. M. Ankowski,
arXiv:1601.06169 [hep-ph]) is related to the first class current
contribution but not to the second class one as has been pointed out
by C. Giunti, arXiv: 1602.00215 [hep--ph].
\end{abstract}


\date{\today}

\maketitle

In the paper Leitner {\it et al.} \cite{Leitner2006} the matrix
element of the transition $n \to p$, caused by the charged vector
current $V^{(+)}_{\mu}(0)$ where $n$ and $p$ are the neutron and
proton, has been written in the following form
\begin{eqnarray}\label{eq:1}
\langle p(k_p,\sigma_p)|V^{(+)}_{\mu}(0)|n(k_n,\sigma_n)\rangle =
  \bar{u}_p(k_p,\sigma_p)\Big(\Big(\gamma_{\mu} -
    \frac{q_{\mu}\hat{q}}{q^2}\Big)\,F_1(q^2) + \frac{i\sigma_{\mu\nu}q^{\nu}}{2
      m_N}\,F_2(q^2)\Big) u_n(k_n,\sigma_n),
\end{eqnarray}
where $ \bar{u}_p(k_p,\sigma_p)$ and $u_n(k_n,\sigma_n)$ are the Dirac
bispinor wave functions of the free proton and neutron in the final
and initial state of the transition $n \to p$, $m_N$ is a nucleon
mass. Then, $q = k_p - k_n$ is the momentum transferred, and
$F_1(q^2)$ and $F_2(q^2)$ are the form factors. The second term in
Eq.(\ref{eq:1}) is caused by the weak magnetism. The right-hand-side
(r.h.s.) of Eq.(\ref{eq:1}) vanishes after multiplication by a
momentum transferred $q^{\mu}$, i.e.
\begin{eqnarray}\label{eq:2}
q^{\mu}\langle p(k_p,\sigma_p)|V^{(+)}_{\mu}(0)|n(k_n,\sigma_n)\rangle = 0,
\end{eqnarray}
even for $m_p \neq m_n$, where $m_p$ and $m_n$ are masses of the
proton and neutron, respectively. The vanishing of the r.h.s. of the
matrix element of the transition $n \to p$ testifies local
conservation of the charged vector current $V^{(+)}_{\mu}(x)$,
i.e. $\partial^{\mu}V^{(+)}_{\mu}(x) = 0$. In other words this
confirms the hypothesis of conservation of the vector current or the
CVC hypothesis by Feynman and Gell-Mann \cite{Feynman1958}. 

The conservation of the charged vector current even for different
masses of the proton and neutron is reached by the contribution of the
phenomenological term $(- q_{\mu}\hat{q}/q^2)\,F_1(q^2)$. Leitner {\it
  et al.}  \cite{Leitner2006} have introduced this term for the
analysis of quasi--elastic scattering $\nu_{\ell} + n \to p + \ell^-$,
where $\nu_{\ell}$ and $\ell^-$ are a neutrino and a lepton with a
lepton flavour $\ell = e,\mu$ and so on.

After Leitner {\it et al.} \cite{Leitner2006} the term $(-
q_{\mu}\hat{q}/q^2)\,F_1(q^2)$ has been used by Ankowski
\cite{Ankowski2016} in the matrix element of the transition $p \to n$
for the calculation of the cross section for the inverse
$\beta$--decay $\bar{\nu}_e + p \to n + e^+$. Such an approach to the
analysis of the inverse $\beta$--decay has been criticized by Giunti
\cite{Giunti2016}. As has been pointed out by Giunti, the vector part
of the matrix element of the transition $p \to n$ should be written in
the following general form 
\begin{eqnarray}\label{eq:3}
\langle n(k_n,\sigma_n)|V^{(-)}_{\mu}(0)|p(k_p,\sigma_p)\rangle =
\bar{u}_n(k_n,\sigma_n)\Big(\gamma_{\mu} \,F_1(q^2) +
\frac{i\sigma_{\mu\nu}q^{\nu}}{2 m_N}\,F_2(q^2) +
\frac{q_{\mu}}{m_N}\,F_3(q^2)\Big) u_p(k_p,\sigma_p),
\end{eqnarray}
where $q = k_n - k_p$ and $F_j(q^2)$ for $j = 1,2,3$ are a momentum
transferred and form factors as functions of $q^2$,
respectively. Multiplying both sides of Eq.(\ref{eq:3}) by a momentum
transferred $q^{\mu}$ and using the Dirac equation for the free proton
and neutron, i.e. replacing $\bar{u}_n(k_n,\sigma_n)\hat{q}
u_p(k_p,\sigma_p)\,F_1(q^2) $ by $(m_n -
m_p)\,F_1(q^2)\bar{u}_n(k_n,\sigma_n) u_p(k_p,\sigma_p)$, Giunti has
found that for $m_p \neq m_n$ the charged vector current conserves
only if $F_3(q^2)$ is related to $F_1(q^2)$ by
\begin{eqnarray}\label{eq:4}
F_3(q^2) = - \frac{(m_n - m_p)}{q^2}\,m_N F_1(q^2).
\end{eqnarray}
Then, Giunti has pointed out that since the term $q_{\mu}F_3(q^2)$ in
the matrix element Eq.(\ref{eq:3}) is induced by the second class
current \cite{Weinberg1958} and, correspondingly, should be prohibited
\cite{Weinberg1958}, the relation Eq.(\ref{eq:4}) should not exist. As
a consequence of this assertion, since the r.h.s. of Eq.(\ref{eq:4}),
taken between the Dirac bispinor wave functions of the neutron and
proton, can be transcribed into the form $(- \hat{q}/q^2)\,m_N
F_1(q^2)$, Giunti has claimed that the term $(-
q_{\mu}\hat{q}/q^2)\,F_1(q^2)$ in the matrix element of the transition
$p \to n$ should be prohibited as induced by the second class current.

We agree with the assertion that the relation Eq.(\ref{eq:4}) should
not exist at all, and the term $q_{\mu} F_3(q^2)$ is the contribution
of the second class current. However, it does not entail a suppression
of the term $(- q_{\mu}\hat{q}/q^2)\,F_1(q^2)$ in the matrix elements
of the transitions $p \to n$ and $n \to p$. The point is that the
terms $q_{\mu} F_3(q^2)$ and $( - q_{\mu}\hat{q}/q^2)\,F_1(q^2)$ are
induced by currents of different classes. That is why for the most
general form of the matrix element of the transition $p \to n$ one may
propose the following expression
\begin{eqnarray}\label{eq:5}
\langle n(k_n,\sigma_n)|V^{(-)}_{\mu}(0)|p(k_p,\sigma_p)\rangle =
\bar{u}_n(k_n,\sigma_n)\Big(\gamma_{\mu} \,F_1(q^2) +
\frac{i\sigma_{\mu\nu}q^{\nu}}{2 m_N}\,F_2(q^2) +
\frac{q_{\mu}\hat{q}}{m^2_N}\,F_4(q^2) +
\frac{q_{\mu}}{m_N}\,F_3(q^2) \Big) u_p(k_p,\sigma_p).
\end{eqnarray}
The first three terms in the r.h.s. of Eq.(\ref{eq:5}) are the
contributions of the first class currents, whereas the fourth term is
caused by the second class one. As a consequence of conservation of
the charged vector current even for $m_p \neq m_n$ one gets $F_3(q^2)
= 0$ and $F_4(q^2) = - (m^2_N/q^2)\,F_1(q^2)$, respectively, without
turning to the Dirac equations for the free proton and neutron. Let us
now show that the term $(q_{\mu}\hat{q}/m^2_N)\,F_4(q^2)$ is the
contribution of the first class current.

As has been pointed out by Weinberg \cite{Weinberg1958}, the hadronic
weak currents can be classified according to their properties under
the $G$--parity transformation \cite{Lee1956}, where $G = C e^{\,i \pi
  I_2}$ is a product of the charge conjugation $C$ and rotation in
isospin space around the second axis $e^{\,i \pi I_2}$, and $I_2$ is
the isospin operator \cite{Lee1956}. For the nucleon it is equal to
$I_2 = \frac{1}{2}\,\tau_2$, where $\tau_2$ is the Pauli isospin
matrix \cite{Weinberg1958}. Under the $G$--parity transformation the
nucleon field operator transforms as follows \cite{Weinberg1958}
\begin{eqnarray}\label{eq:6}
GN(x)G^{-1} = N^G(x) = i\tau_2 N^C(x) = i\tau_2 C \bar{N}^T
\quad,\quad G\bar{N}(x)G^{-1} = \overline{N^G}(x) = N^T(x) C (-
i)\tau_2,
\end{eqnarray}
where $C = i\gamma^0\gamma^2$ is the charge conjugation matrix with
the property $\gamma^0 C \gamma^0 = - C$ , $T$ is a transposition, and
$\bar{N}(x) = N^{\dagger}(x)\gamma^0$ \cite{Itzykson1980}. Then,
$N(x)$ is the nucleon isospin doublet with components $(p(x), n(x))$,
where $p(x)$ and $n(x)$ are the field operators of the proton and
neutron, respectively, and $\overline{N^G}(x) =
N^{G\dagger}(x)\gamma^0$. In turn, $N^G(x) = i\tau_2 N^C$ is the
antinucleon isospin doublet with the components $(n^C(x), - p^C(x))$,
where $n^C(x)$ and $p^C(x)$ are the field operators of the antineutron
and antiproton, respectively. Below we analyse the $G$--parity
transformation properties of the following currents i)
$\bar{N}(x)\gamma_{\mu}\tau^{(\mp)} N(x)$, ii)
$\partial^{\nu}(\bar{N}(x)\sigma_{\mu\nu}\tau^{(\mp)} N(x))$, iii)
$\partial_{\mu}\partial^{\nu}(\bar{N}(x)\gamma_{\nu}\tau^{(\mp)}
N(x))$ and iv) $\partial_{\mu}(\bar{N}(x)\tau^{(\mp)} N(x))$, inducing
the terms with the Lorentz structures i) $\gamma_{\mu}$, ii)
$i\sigma_{\mu\nu}q^{\nu}$, iii) $q_{\mu} \hat{q}$ and iv) $q_{\mu}$,
respectively, in the matrix elements of the transition $p \to n$ (see
Eq.(\ref{eq:5})), where $\tau^{(-)} = (\tau_1 - i\tau_2)/2$ and
$\tau^{(+)} = (\tau_1 + i \tau_2)/2$ stand for the transition $p \to
n$ and $p \to n$, respectively, and $\vec{\tau} = (\tau_1, \tau_2,
\tau_3)$ are the isospin Pauli matrices.

\subsection{Properties of the current $\bar{N}(x)\gamma_{\mu}\tau^{(\mp)}
 N(x)$ under the $G$--parity transformation}

The current $\bar{N}(x)\gamma_{\mu}\tau^{(\mp)} N(x)$ possesses the
 following property under the $G$--parity transformation:
\begin{eqnarray}\label{eq:7}
\hspace{-0.3in}&&\bar{N}(x)\gamma_{\mu}\tau^{(\mp)} N(x)
\stackrel{G}{\longrightarrow} G\bar{N}(x)\gamma_{\mu}\tau^{(\mp)} N(x)
G^{-1} = \overline{N^G}(x)\gamma_{\mu}\tau^{(\mp)} N^G(x) = N^T(x)C(-
i) \tau_2 \gamma_{\mu} \tau^{(\mp)} i\tau_2 C \bar{N}^T(x)
=\nonumber\\
\hspace{-0.3in}&&= - N^T(x)C\gamma_{\mu}C \tau^{(\pm)} \bar{N}^T(x) =
- N^T(x) \gamma^T_{\mu}\tau^{(\pm)} \bar{N}^T(x) =
\bar{N}(x)\gamma_{\mu}\tau^{(\pm)T} N(x) =
\bar{N}(x)\gamma_{\mu}\tau^{(\mp)} N(x).
\end{eqnarray}
This is the $G$--parity transformation property of the first class
current \cite{Weinberg1958}, where we have used the relations $i
\tau_2 \tau^{(\mp)} i\tau_2 = \tau^{(\pm)}$, $\tau^{(\pm)T} =
\tau^{(\mp)}$, $C\gamma_{\mu}C = \gamma^T_{\mu}$ and $N^T(x)
\bar{N}^T(x) = - \bar{N}(x) N(x)$ \cite{Itzykson1980}. The current
$\bar{N}(x)\gamma_{\mu}\tau^{(\mp)} N(x)$ is responsible for the
Lorentz structure $\gamma_{\mu}$ in Eq.(\ref{eq:5}).

\subsection{Properties of the current $\partial^{\nu}(\bar{N}(x) 
\sigma_{\mu\nu}\tau^{(\mp)} N(x))$ under the $G$--parity
transformation}

Under the $G$--parity transformation the current
$\partial_{\nu}(\bar{N}(x)\sigma_{\mu\nu}\tau^{(\mp)} N(x))$
transforms as follows
\begin{eqnarray}\label{eq:8}
\hspace{-0.3in}&&
\partial^{\nu}\Big(\bar{N}(x)\sigma_{\mu\nu}\tau^{(\mp)} N(x)\Big)
\stackrel{G}{\longrightarrow}
G\partial^{\nu}\Big(\bar{N}(x)\sigma_{\mu\nu}\tau^{(\mp)} N(x)\Big)
G^{-1} = \partial^{\nu}\Big(\overline{N^G}(x) \sigma_{\mu\nu}
\tau^{(\mp)} N^G(x)\Big) = \nonumber\\
\hspace{-0.3in}&&= \partial^{\nu}\Big(N^T(x)C (- i) \tau_2
\sigma_{\mu\nu}\tau^{(\mp)} i\tau_2 C \bar{N}^T(x)\Big) =
-\partial^{\nu}\Big(N^T(x)C \sigma_{\mu\nu}C \tau^{(\pm)}
\bar{N}^T(x)\Big) = \nonumber\\
\hspace{-0.3in}&&= -\partial^{\nu}\Big( N^T(x)\sigma^T_{\mu\nu}
\tau^{(\pm)} \bar{N}^T(x) =
\partial^{\nu}\Big(\bar{N}(x)\sigma_{\mu\nu} \tau^{(\pm)T} N(x)\Big) =
\partial^{\nu}\Big(\bar{N}(x)\sigma_{\mu\nu} \tau^{(\mp)} N(x)\Big).
\end{eqnarray}
Hence, $\partial_{\nu}(\bar{N}(x)\sigma_{\mu\nu}\tau^{(\mp)} N(x))$ is
the first class current \cite{Weinberg1958}. For the derivation of
Eq.(\ref{eq:8}) we have used the relation $C\sigma_{\mu\nu} C =
\sigma^T_{\mu\nu}$ \cite{Itzykson1980}. The current
$\partial_{\nu}(\bar{N}(x)\sigma_{\mu\nu}\tau^{(\mp)} N(x))$ is
responsible for the Lorentz structure $i\sigma_{\mu\nu}q^{\nu}$ in
Eq.(\ref{eq:5}).

\subsection{Properties of the current $\partial_{\mu}\partial^{\nu} 
 (\bar{N}(x)\gamma_{\nu}\tau^{(\mp)} N(x))$ under the $G$--parity
  transformation}

Making the $G$--parity transformation we get
\begin{eqnarray}\label{eq:9}
\hspace{-0.3in}&&\partial_{\mu}\partial^{\nu}\Big(\bar{N}(x)
\gamma_{\nu}\tau^{(\mp)} N(x)\Big) \stackrel{G}{\longrightarrow}
G \partial_{\mu}\partial^{\nu}\Big(\bar{N}(x)\gamma_{\nu}\tau^{(\mp)}
N(x)\Big) G^{-1} =
\partial_{\mu}\partial^{\nu}\Big(\overline{N^G}(x))\gamma_{\nu}
\tau^{(\mp)} N^G(x)\Big) = \nonumber\\
\hspace{-0.3in}&&= \partial_{\mu}\partial^{\nu}\Big(N^T(x)C (- i)
\tau_2\gamma_{\nu} \tau^{(\mp)} i\tau_2 C \bar{N}^T(x)\Big) = -
\partial_{\mu}\partial^{\nu}\Big(N^T(x) \gamma^T_{\nu} \tau^{(\pm)}
\bar{N}^T(x)\Big) =
\partial_{\mu}\partial^{\nu}\Big(\bar{N}(x)\gamma_{\nu}\tau^{(\pm)T}
N(x)\Big) =\nonumber\\
\hspace{-0.3in}&&= 
\partial_{\mu}\partial^{\nu}\Big(\bar{N}(x)\gamma_{\nu}\tau^{(\mp)}
N(x)\Big).
\end{eqnarray}
This testifies that $\partial_{\mu}\partial^{\nu} (\bar{N}(x)
\gamma_{\nu}\tau^{(\mp)} N(x))$ is the first class current, which is
responsible for the Lorentz structure $q_{\mu} \hat{q}$ in
Eq.(\ref{eq:5}).

\subsection{Properties of the current  $\partial_{\mu}(\bar{N}(x) 
\tau^{(\mp)} N(x))$ under the $G$--parity transformation}

Now we are turning to the analysis of the properties of the current
$\partial_{\mu}(\bar{N}(x)\tau^{(\mp)} N(x))$ under the $G$--parity
transformation. Making the $G$--parity transformation we obtain
\begin{eqnarray}\label{eq:10}
\hspace{-0.3in}&&\partial_{\mu}\Big(\bar{N}(x)\tau^{(\mp)} N(x)\Big)
\stackrel{G}{\longrightarrow}
G \partial_{\mu}\Big(\bar{N}(x)\tau^{(\mp)} N(x)\Big) G^{-1} =
\partial_{\mu}\Big(\overline{N^G}(x)\tau^{(\mp)} N^G(x)\Big) =\nonumber\\
\hspace{-0.3in}&&= - \partial_{\mu}\Big(N^T(x)C i \tau_2 \tau^{(\mp)}
i\tau_2 C \bar{N}^T(x)\Big) - \partial_{\mu}\Big(N^T(x) C^2
\tau^{(\pm)} \bar{N}^T(x)\Big) = - \partial_{\mu}\Big(N^T(x) (-1)
\tau^{(\pm)} \bar{N}^T(x)\Big) = \nonumber\\
\hspace{-0.3in}&&= - \partial_{\mu}\Big(\bar{N}(x)\tau^{(\pm)T}
N(x)\Big) = - \partial_{\mu}\Big(\bar{N}(x)\tau^{(\mp)} N(x)\Big).
\end{eqnarray}
 Hence, the current $\partial_{\mu}(\bar{N}(x)\tau^{(\mp)} N(x))$
 belongs to the second class \cite{Weinberg1958}.  It is responsible
 for the Lorentz structure $q_{\mu}$ in Eq.(\ref{eq:5}). For the
 derivation of Eq.(\ref{eq:10}) we have used the property of the
 charge conjugation matrix $C^2 = - 1$ \cite{Itzykson1980}.

Thus, we have shown that the terms with the Lorentz structures
$q_{\mu}$ and $ q_{\mu} \hat{q}$ in the matrix element of the
transition $p \to n$ are induced by the second and first class
currents, respectively. This means that the form factor $F_3(q^2)$
cannot be expressed in terms of the form factor $F_1(q^2)$, which is
caused by the first class current, by simply multiplying
Eq.(\ref{eq:3}) or Eq.(\ref{eq:5}) by a momentum transferred
$q^{\mu}$. Of course, it may vanish as it is required by conservation
of the charged vector current, but it does not lead to the suppression
of the term $(q_{\mu} \hat{q}/m^2_N)\,F_4(q^2)$. As a result, the form
factor $F_4(q^2)$ is able to be equal to $F_4(q^2) = -
(m^2_N/q^2)\,F_1(q^2)$ in order to maintain conservation of the
charged vector current even for $m_p \neq m_n$.

Our analysis concerns fully the matrix element (see Eq.(\ref{eq:1}))
of the transition $n \to p$ in quasi--elastic scattering $\nu_{\ell} +
n \to p + \ell^-$ \cite{Leitner2006} and in the neutron
$\beta^-$--decay $n \to p + e^- + \bar{\nu}_e$
\cite{Ivanov2017a,Ivanov2017b} (see also neutrino production of the
$N(1440)$--resonance (Roper resonance) on nucleon \cite{Paschos2006}).

This work was supported by the Austrian ``Fonds zur F\"orderung der
Wissenschaftlichen Forschung'' (FWF) under Contracts I689-N16,
P26781-N20 and P26636-N20 and ``Deutsche F\"orderungsgemeinschaft''
(DFG) AB 128/5-2.

\end{document}